\documentclass[10pt,english]{article}
\usepackage{babel}
\usepackage{verbatim,graphicx} 
\usepackage[T1]{fontenc}
\usepackage{amsmath}

\def\u#1{{\mathbf #1}}

\def\ub#1{{\mathbf #1}}
\def\ub#1{{\boldsymbol #1}}

\def\be{\begin{equation}}
\def\ee{\end{equation}}
\def\bea{\begin{eqnarray}}
\def\eea{\end{eqnarray}}
\def\bd{\begin{displaymath}}
\def\ed{\end{displaymath}}
\def\RR{\mathcal{R}}
\newcommand{\cpvg}[1]{%
  \mbox{ }\makebox[0pt]{$-$}\makebox[0pt]{$\displaystyle \int$}_{#1}\mbox{ }}
\newcommand{\cpvgs}[1]{%
  \mbox{ }\makebox[0pt]{$-$}\makebox[0pt]{$\int$}_{#1}\mbox{ }}

\begin{document}

\begin{center}
\begin{Large}
\textsc{A Simple and Efficient Regularization Method for 3D BEM:}\\
\textsc{Application to Frequency-Domain Elastodynamics}
\end{Large}
\vspace{0.5cm}

Patrick Dangla$^*$, Jean-François Semblat$^{**,}$\footnote{corresponding author: \texttt{semblat@lcpc.fr}}, Haihong Xiao$^{**}$, Nicolas Delépine$^{**}$\\
\vspace{0.2cm}
\small{\em $^*$ LCPC/LMSGC, 2 allée Kepler, 77420 Champs-sur-Marne, France}\\
\vspace{0.1cm}
\small{\em $^{**}$ LCPC, 58 bd Lefebvre, 75732 Paris Cedex 15, France}\\
\vspace{0.2cm}
\end{center}




\begin{abstract}
An efficient and easy-to-implement method is proposed to regularize integral equations in the 3D Boundary Element Method. The method takes advantage of an assumed three-noded triangle discretization of the boundary surfaces. The method is based on the derivation of analytical expressions of singular integrals. To demonstrate the accuracy of the method, three elastodynamic problems are numerically worked out in frequency domain: cavity under harmonic pressure, diffraction of a plane wave by a spherical cavity, amplification of seismic waves in a semi-spherical alluvial basin (the second one is also investigated in time domain). The numerical results are compared to (semi-) analytical solutions; a close agreement is found for all problems showing the very good accuracy of the proposed method.
\end{abstract}


\begin{twocolumn}
\indent
\begin{center}
{\large{Regularization of Boundary Integral Equations}}
\end{center}

In contrast to other discretization methods, the Boun\-dary Element Method involves singular integrals. According to the power, there are three kinds of singularities that depend on whether integrability is defined (\emph{i}) in the ordinary Riemann sense (weak singularity), (\emph{ii})~in the Cauchy principal value sense (strong singularity) or (\emph{iii}) in the Hadamard finite part sense (hyper-singularity). Strong singular integrals appear in the Ordinary Boundary Integral Equations (OBIE) while the Derivative Boundary Integral Equations (DBIE) involve both the strong and hyper-singular integrals. Strong and hyper-singular integrals have to be converted to regular ones in the regularization of the BEM formulations (Tanaka {\em et al.}, 1994; Sladek and Sladek, 1996, 1998). Strictly speaking, weak singularity is not treated by regularization. However, from the point of view of numerical integrations, one should devote a great attention to the evaluation of these integrals because standard integration quadratures fail in accuracy (Lachat and Watson, 1976; Sladek {\em et al.}, 1997, 2001; Manolis and Beskos, 1988). Therefore each type of singularity has to be treated by appropriate techniques. Most of the researches has dealt with strong (Bonnet and Bui, 1993) and hyper-singularities. Some methods have been proposed in the literature to treat these singular integrals (Sladek and Sladek, 1996; Niu and Zhou, 2004; Guiggiani and Gigante, 1990; Guiggiani {\em et al.}, 1992; Chen and Hong, 1999; Bonnet, 1999; Bui {\em et al.}, 1985; Bonnet and Xiao, 1995; Aubry and Clouteau, 1991; Xiao, 1994; Guiggiani, 1994). It is noticed that the regularization can be performed either before or after the discretization, i.e. in the global or local (intrinsic) coordinate space, as observed in some papers mentioned above. A comprehensive review of BEM in dynamic analysis has been proposed by Beskos (Beskos, 1997).\\

In this paper, the regularization is performed in the global coordinate space after the discretization of the geometry. Herein, only strong and weak singularities of the ordinary boundary integral equations are dealt with. The method takes advantage of an assumed three-noded triangle element for the discretization of three-dimensional problems. Thanks to this simple shaped element, one can performed analytical derivations of the Cauchy principal value of the singular integrals. Such an approach has been previously applied for 2D elastodynamic problems in (Dangla, 1988, 1990). To the author's best knowledge, this method has not been used in any 3D analysis. This paper addresses this issue by summarising the theoretical background of the method. Afterwards, the efficiency and accuracy of the regularization method is analysed in 3D elastodynamics.\\

\begin{center}
{\large{Numerical Modeling in Elastodynamics}}
\end{center}

To analyze problems in 3D elastodynamics, various numerical methods are available:
\begin{itemize}
  \item the {\em finite element method} which is efficient to deal with complex geometries and numerous heterogene\-ities (Chammas {\em et al.}, 2003), even for inelastic constitutive models (Bonilla, 2000). It has nevertheless several drawbacks such as numerical dispersion (and damping) (Ihlenburg and  Babu\-\v{s}ka, 1995; Semblat and Brioist, 2000) and (consequently) numerical cost in 3D elastodynamics,
  \item the {\em finite difference method} which is very accurate in elastodynamics but is mainly adapted to simple geometries and linear constitutive models (Frankel and Vidale, 1992; Moczo {\em et al.}, 2002; Virieux, 1986)
  \item the {\em boundary element method} which allows a very good description of the radiation conditions but is preferably dedicated to weak heterogeneities and linear constitutive models (Banerjee {\em et al.}, 1988; Beskos, 1997; Beskos {\em et al.}, 1986; Bonnet, 1999; Dangla, 1988; Sánchez-Sesma and Luzón, 1995; Yokoi, 2003)
  \item the {\em spectral element method} which has been increasingly considered to analyse 2D/3D wave propagation in linear media (Faccioli {\em et al.}, 1996; Komatitsch and Vilotte, 1998)
  \item the {\em Aki-Larner method} which takes advantage of the frequency-wave\-number decomposition but is limited to simple geometries (Aki and Larner, 1970; Bouchon {\em et al.}, 1989)
  \item {\em series expansions of wave functions} which give a semi-analytical estimation of the scattered wavefield for simple geometries (Moeen-Vaziri and Trifunac, 1985; Sán\-chez-Sesma, 1983)
\end{itemize}
Each method has specific advantages and drawbacks. It is consequently often more interesting to combine two methods to take advantage of their peculiarities. One of the most common method in elastodynamics is to couple FEM and BEM allowing an accurate description of the near field (FEM model including complex geometries, heterogeneities and constitutive behaviours) and a reliable estimation of the far-field (BEM model involving radiation conditions).\\

\begin{center}
{\large{Integral Equations}}
\end{center}

This paper is limited to isotropic elastodynamics for time-harmonic problems of circular frequency $\omega$. For any given body force distribution $F_i(\u{x})$ over $\Omega$, the governing equations which must be verified by any displacement and stress fields, $u_i(\u{x})$ and $\sigma_{ij}(\u{x})$, take the following form:
\be
\sigma_{ij} = \lambda u_{k,k}\delta_{ij} + \mu(u_{i,j} + u_{j,i}) 
\label{eq:elasticity}
\ee
\be
\sigma_{ij,j} + \rho \omega^2 u_i + F_i = 0
\label{eq:balance_eq}
\ee
The fundamental solutions, in time-harmonic elastodynamics, are defined by a force of unit amplitude applied at a fixed point $\u{y}$ and in a fixed coordinate direction $k$: $F_i(\u{x}) = \delta(\u{x}-\u{y})\delta_{ik}$. For infinite body the fundamental solution, denoted by $u_i(\u{x})=U_i^k(\u{x},\u{y};\omega)$, is known as the Helmholtz fundamental solution and is given by (Eringen and Suhubi, 1975):

\begin{eqnarray}
U_i^k(\u{x},\u{y};\omega) = \frac{1}{4\pi \mu} \left[\frac{1}{k_S^2}\frac{\partial^2}{\partial x_i\partial x_k}\left (\frac{e^{ik_Sr}}{r}-\frac{e^{ik_Pr}}{r} \right)\right. \label{eq:Uik_helmholtz}\\
\nonumber\left.+\frac{e^{ik_Sr}}{r}\delta_{ik}\right]
\end{eqnarray}

where $r^2=(\u{x}-\u{y})^2$ and where $k_P=\omega\sqrt{\rho/(\lambda+2\mu)}$ and $k_S=\omega\sqrt{\rho/\mu}$ are the longitudinal and transversal wave numbers respectively.
The stress tensor associated with $U_i^k(\u{x},\u{y};\omega)$, defined by (\ref{eq:elasticity}), is denoted by $\Sigma_{ij}^k(\u{x},\u{y};\omega)$ while the stress vector applied to the surface boundary of $\Omega$ is $T_i^k(\u{x},\u{y};\omega) = \Sigma_{ij}^k(\u{x},\u{y};\omega)n_j$.

For sake of simplicity let us assume no body force from now on. Application of the Maxwell-Betti reciprocity theorem leads to the following displacement integral representation at point $\u{y} \in \RR^3$ (Bonnet, 1999): 
\begin{eqnarray}
\kappa(\u{y}) u_k(\u{y}) = \int_{\partial\Omega} \left[ t_i(\u{x}) U_i^k(\u{x},\u{y};\omega) \right.~~~~~~~~~\label{eq:rep-formula}\\
\nonumber\left. - u_i(\u{x})T_i^k(\u{x},\u{y};\omega) \right]\,dS_x
\end{eqnarray}
where $\kappa = 1$ ($\u{y} \in \Omega$) or $\kappa = 0$ ($\u{y} \not\in \Omega$).

Let $\u{y}$ denote a fixed point on the boundary surface $\partial\Omega$. For a given small $\varepsilon > 0$, introduce a spherical shaped neighbourhood $v_\varepsilon(\u{y})$ of $\u{y}$, called an exclusion neighbourhood (Guiggiani {\em et al.}, 1992). The domain $\Omega_{\varepsilon}(\u{y}) = \Omega - v_\varepsilon(\u{y})$ obtained by removing $v_\varepsilon(\u{y})$ from $\Omega$ is such that the point $\u{y}$ is exterior to $\Omega_{\varepsilon}(\u{y})$. Its boundary is $\partial\Omega_{\varepsilon} = (\partial\Omega - e_{\varepsilon}) + s_{\varepsilon}$, where $e_{\varepsilon} = \partial\Omega\cap v_\varepsilon$, $s_{\varepsilon} = \Omega\cap \partial v_\varepsilon$. The classical form of the integral equation consists in taking the limit $\varepsilon \rightarrow 0$ in the representation formula (\ref{eq:rep-formula}) taken for the domain $\Omega_{\varepsilon}$. The limiting expression thus obtained is known as the \emph{Somigliana identity}:
\begin{eqnarray}
C_i^k(\u{y}) u_i(\u{y}) = \cpvg{\partial\Omega}
\left[ t_i(\u{x})U_i^k(\u{x},\u{y};\omega)\right.~~~~~~~~~\label{eq:ei}\\
\nonumber\left. - u_i(\u{x})T_i^k(\u{x},\u{y};\omega) \right] \,dS_x
\end{eqnarray}
The notation $\cpvgs{~}$ stands for the \emph{Cauchy principal value} of a singular integral, i.e. the limit:
\begin{equation}
\cpvg{\partial\Omega}(\cdot) = \lim_{\varepsilon\rightarrow 0}\int_{(\partial\Omega-e_{\varepsilon})} (\cdot)
\end{equation}
The \emph{free term} $C_i^k(\u{y})$ appearing in (\ref{eq:ei}), is defined by:
\begin{equation}
C_i^k(\u{y})=\lim_{\varepsilon\rightarrow 0} \int_{s_\varepsilon} T_i^k(\u{x},\u{y};\omega)\,dS_x
\label{eq:cij}
\end{equation}
It is found to be equal to $1/2\delta_{ik}$ when $\Omega$ is smooth at $\u{y}$.\\

\begin{center}
{\large{Discretization and Regularization Principle}}
\end{center}

The boundary surface $\partial\Omega$ and the boundary variables $(u_i,t_i)$ are discretized by using three-noded triangular elements. A finite set of equations is generated by enforcing equation (\ref{eq:ei}) at the nodes of the surface mesh (collocation method). Thus the boundary surface consists of the set of $N$ boundary surface elements $E_e$: $\partial\Omega = \{E_e, e=1...N\}$. The integral appearing in (\ref{eq:ei}) then assumes the form of a sum of $N$ element integrals:
\begin{equation}
\cpvg{\partial\Omega}(\cdot) = \sum_{e=1}^N \cpvg{E_e}(\cdot) 
\label{eq:discret-int} 
\end{equation}
The numerical evaluation of non singular element integrals that appear in (\ref{eq:discret-int}) is usually based, like in finite element methods, on Gaussian quadrature formulas. The approximate value of an element integral can be given formally by:
\begin{equation}
\int_{E_e} f(\u{x})\,dS_x \approx \stackrel{\mbox{\scriptsize Nu}}{\int_{E_e}} f(\u{x})\,dS_x = \sum_{i=1}^n w_if(\u{x}_i)
\label{eq:num-int}
\end{equation}
where $\u{x}_i$ and $w_i$ are the coordinates and weights of the \emph{Gauss points}. The notation $\stackrel{\mbox{\scriptsize Nu}}{\int}$ stands for the numerical approximation of integrals. This special notation has been adopted to emphasize that in case of singular integral $\cpvgs{~} \neq \stackrel{\mbox{\scriptsize Nu}}{\int}$.

Since some of element integrals are singular, a straighforward evaluation of (\ref{eq:ei}) based on Gaussian quadrature formulas will inevitably lead to some significant error. To correct this error, a new term $R^k(\u{y})$ must be introduced in the numerical evaluation of (\ref{eq:ei}):
\begin{eqnarray}
C_i^k(\u{y}) u_i(\u{y}) =  R^k(\u{y}) + \stackrel{\mbox{\scriptsize Nu}}{\int_{\partial\Omega}}
\left[ t_i(\u{x})U_i^k(\u{x},\u{y};\omega) \right.~~~~~~~~\label{eq:ei-reg}\\
\nonumber\left. -  u_i(\u{x})T_i^k(\u{x},\u{y};\omega) \right] \,dS_x
\end{eqnarray}

The regularization method proposed in this paper consists in deriving analytically the correction term by taking advantage of the simple shape of the triangle elements. To do so, let us introduce the Kelvin's fundamental solution:
\begin{equation}
U_i^k(\u{x},\u{y}) = \frac{1}{16\pi\mu(1-\nu) r} \left (r_{,i}r_{,k} + (3-4\nu) \delta_{ik} \right )
\label{eq:Uik_kelvin}
\end{equation}
and note  $\Sigma_{ij}^k(\u{x},\u{y})$ the stress tensor associated with solution $U_i^k(\u{x},\u{y})$. It is noticed that the Helmholtz and Kelvin solutions have identical singularities:
\begin{equation}
\begin{array}{l}
(\u{x}\rightarrow\u{y})\\
(\forall \omega > 0)
\end{array}
\left\{
\begin{array}{l}
U_i^k(\u{x},\u{y};\omega) - U_i^k(\u{x},\u{y}) = O(1) \\
\Sigma_{ij}^k(\u{x},\u{y};\omega) - \Sigma_{ij}^k(\u{x},\u{y}) = O(1)
\end{array}
\right.
\end{equation}

Thanks to this property the correction term $R^k(\u{y})$ only needs to involve the Kelvin fundamental solutions. For a given point $\u{y} \in \partial\Omega$, introduce the index subset $I(\u{y}) = \{e \in [1,N], \u{y} \in E_e\}$ such that the integral over $E_e$ is singular for $e \in I(\u{y})$ and non singular for $e \notin I(\u{y})$. Introduce $\partial\Omega_{y} = \{E_e, e \in I(\u{y})\}$ the set of the neighborhood elements of $\u{y}$. Thus the correction term can be formulated in the following form:

\begin{eqnarray}
R^k(\u{y})=~~~~~~~~~~~~~~~~~~~~~~~~~~~~~~~~~~~~~~~~~~~~~~~~~~~~~~~\label{eq:corr}\\
\nonumber \left( \cpvg{\partial\Omega_y} U_i^k(\u{x},\u{y}) dS_x -\stackrel{\mbox{\scriptsize Nu}}{\int_{\partial\Omega_y}} U_i^k(\u{x},\u{y}) dS_x \right) t_i(\u{y})\\
\nonumber -\left( \cpvg{\partial\Omega_y} T_i^k(\u{x},\u{y}) dS_x
-\stackrel{\mbox{\scriptsize Nu}}{\int_{\partial\Omega_y}} T_i^k(\u{x},\u{y}) dS_x \right) u_i(\u{y})
\end{eqnarray}

It can be noticed that formulation~(\ref{eq:corr}) is independent of the interpolation order since $t_i(\u{y})$ and $u_i(\u{y})$ only need to be evaluated at point $\u{y}$. Taking advantage of the simple shape of the three-noded triangle elements, we can derive analytical expressions of the singular integrals appearing in~(\ref{eq:corr}).\\

In particular, it can be shown that they are the sum of elementary contributions involving elements of $I(\u{y})$:
\begin{eqnarray}
I_i^k(\u{y}) = \cpvg{\partial\Omega_y}  T_i^k(\u{x},\u{y}) dS_x = \!\!\sum_{e \in I(\u{y})} I_i^k(\u{y};E_e)
\label{eq:I1}  \\
J_i^k(\u{y}) = \cpvg{\partial\Omega_y} U_i^k(\u{x},\u{y}) dS_x = \!\!\sum_{e \in I(\u{y})} J_i^k(\u{y};E_e)
\label{eq:J1}
\end{eqnarray}
The analytical derivations of $I_i^k(\u{y};E_e)$ and $J_i^k(\u{y};E_e)$ are proposed in the appendix (equations~(\ref{eq:I1_1}) to (\ref{eq:j1_2})).

In a similar manner, the free term involves the Kelvin fundamental solutions and can be assumed in the form of a sum of \emph{free term elements } involving elements of $I(\u{y})$:
\begin{equation}
C_i^k(\u{y}) = \lim_{\varepsilon\rightarrow 0} \int_{s_\varepsilon} T_i^k(\u{x},\u{y})\,dS_x \label{eq:discret-cij} = \!\!\sum_{e \in I(\u{y})} C_i^k(\u{y};E_e)
\end{equation}
where the exact derivation of $C_i^k(\u{y};E_e)$ is given in the appendix (equations~(\ref{eq:cij_0}) to (\ref{eq:cij_7})).

Since the method of derivation of the correction term $R^k(\u{x},\u{y})$ is now established, the formulation (\ref{eq:ei-reg}) can be considered as the regularized form of the initial integral equation (\ref{eq:ei}).\\

\begin{center}
{\large{Numerical Implementation}}
\end{center}

Both boundary and unknowns are discretized using three-noded flat triangles and interpolation techniques initially developed for the Finite Element Method. The discretization of the geometry and the unknowns is thus written, respectively, as follows (Bonnet, 1999):
\bea
\u{x}(\ub{\xi})=\sum_{k=1}^3 N_k(\ub{\xi})\u{x}^k~~ &
\displaystyle ~~a(\u{x})=\sum_{k=1}^3 N_k(\ub{\xi})a^k
\eea 
with $\u{x}^k$: the node coordinates, $N_k$: the linear interpolation functions and $a^k$: the {\em nodal values} of the displacement or traction unknowns.\\
Thus, the set of scalar equations resulting from the discretization of equations~(\ref{eq:ei-reg}), enforced at the nodes of the mesh, has the following matrix structure:
\begin{equation}
  \mathbf{[A] \{u\} + [B] \{t\} = 0}
\label{eq:eAB}
\end{equation}
where $\mathbf{[A]}$ and $\mathbf{[B]}$ are fully populated non symmetric matrices. $\mathbf{\{u\},\{t\}}$ are the ``vectors'' containing, respectively, the nodal values of $u_i(\u{y})$ and $t_i(\u{y})$. The incorporation of the boundary conditions consists in substituting the prescribed nodal values of $(u_i,t_i)$ into $\mathbf{\{u\},\{t\}}$ in Eq.~(\ref{eq:eAB}). The columns of this matrix equation are reordered so as to have a matrix equation of the form:
\begin{equation}
  \mathbf{[K] \{v\} = \{ f \}}
\label{EId}
\end{equation}
where the vector $\mathbf{\{v\}}$ consists of the unknown components of $\mathbf{\{u\},\{t\}}$. The matrix $\mathbf{[K]}$ contains the columns of $\mathbf{[A],[B]}$ associated with those unknown components while the right-hand side $\mathbf{\{f\}}$ results from the multiplication of the known components of $\mathbf{\{u\},\{t\}}$ by the corresponding columns of the matrices $\mathbf{[A],[B]}$. As shown in the following for unbounded media, the right hand side $\mathbf{\{f\}}$ can also involve a contribution due to an incident wavefield. The method has been implemented into the computer code CESAR-LCPC (Humbert {\em et al.}, 2005) of the Laboratoire Central des Ponts et Chaussées (French Public Works Research Laboratory, Paris, France).

\begin{figure}[htp]
\begin{center}
\end{center}
\caption{\em Cavity under harmonic internal pressure: model description.}
\label{fig:cav3dpress_schem}
\end{figure}

\begin{center}
{\large{Validation in Frequency-Domain Elastodynamics}}
\end{center}

{\em Example 1}: Spherical cavity under harmonic internal pressure.\\
\indent{\em Description of the problem and analytical solution.} The first example (figure~\ref{fig:cav3dpress_schem}) concerns a spherical cavity of radius $R$ in a full elastic isotropic space undergoing an internal harmonic pressure. The cavity mesh includes 320 triangular boundary elements (that is 162 nodes) and a special generation process is considered to have a regular triangular mesh of the sphere starting from an icosahedron (Edouard {\em et al.}, 1996) (also see next sections). Using the regularization method proposed herein, we have computed the displacement field around the cavity at various (normalized) frequencies.\\

The validation of the numerical results is made considering the analytical solution in terms of radial displacement $u(r,\omega)$ given by Eringen and Suhubi (1975) as follows:
\begin{equation}
u(r,\omega) = - \frac{P(\omega)R^3(ik_P-1/r).exp\left(ik_P(r-R)\right)}{4 \mu r (1-ik_P R-k_S^2R^2/4)}
\label{eq:dep_cavite}
\end{equation}
where $k_P$ and $k_S$ are the longitudinal and transverse wavenumbers.\\

This equation can be rewritten using normalized distance $\chi=r/R$, normalized frequency $\eta_P=k_PR/\pi$ (that is $\eta_P=2R/\Lambda_P$, $\Lambda_P$ being the longitudinal wavelength) and considering $\displaystyle\upsilon(\chi,\eta_P) = \frac{\mu u(r,\omega)}{P(\omega) R}$. It leads to:
\begin{equation}
\upsilon(\chi,\eta_P) = - \frac{(i\pi\chi\eta_P - 1).exp(i\pi\eta_P (\chi-1))}{(1-i\pi\eta_P-\pi^2 \zeta^2\eta_P^2/4).\chi^2}
\label{eq:dep_cav_adimparam}
\end{equation}
with $\zeta=k_P/k_S=\sqrt{(1-2\nu)/(2-2\nu)}=1/\sqrt{3}$ (that is $\nu$=0.25).

\begin{figure}[htp]
\begin{center}
\end{center}
\caption{\em Normalized radial displacement $\upsilon(\chi,\eta_P)$ (real part) vs  normalized distance $\chi$: comparison between numerical and analytical results for normalized frequencies $\eta_P$=0.01, 0.50, 1.00 and 2.00.}
\label{fig:cav3dpress_valid}
\end{figure}

{\em Comparisons between numerical and analytical results.} In figure~\ref{fig:cav3dpress_valid}, the real part of the normalized radial displacement $\upsilon(\chi,\eta_P)$ defined by equation (\ref{eq:dep_cav_adimparam}) is displayed vs normalized distance $\chi$ for both analytical and numerical solutions at normalized frequencies $\eta_P$=0.01, 0.50, 1.00 and 2.00. For the nearly static case ($\eta_P$=0.01) as well as the fully dynamic cases, the agreement between the numerical results and the analytical ones is very good at all normalized distances. From this first simple example, the reliability and accuracy of the proposed method then appear very good.\\

{\em Efficiency of the regularization method.} We will then investigate the efficiency of the regularization method itself by evaluating the correction term for the same mechanical problem (figure~\ref{fig:cav3dpress_schem}). A non regularized solution is computed by dropping the correction term in (\ref{eq:ei-reg}). In figure~\ref{fig:cav2p3_corrig}, this non regularized solution is compared with both the regularized one and the analytical solution at normalized frequencies $\eta_P$=0.50 (left) and 2.00 (right). These comparisons show that the numerical results without the analytical correction are far from both analytical and corrected numerical solutions. The efficiency of the regularization method then appears very good since the direct computation of the singular integrals leads to very bad results.

\begin{figure}[htp]
\begin{center}
\end{center}
\caption{\em Normalized radial displacement $\upsilon(\chi,\eta_P)$ (real part) vs normalized distance $\chi$: numerical results with and without the analytical correction of the singular integrals for normalized frequencies $\eta_P$=0.50 (left) and 2.00 (right).}
\label{fig:cav2p3_corrig}
\end{figure}

{\em Example 2}: Diffraction of a plane wave by a spherical cavity.\\
\indent{\em Description of the problem and analytical solution.} The second example deals with the diffraction of a plane P-wave ($u_{inc} = U_0\, exp \left[ i(k_P x - \omega t) \right]$ with $U_0$=1), propagating along $x$ axis, by a spherical cavity. The numerical results are firstly computed in frequency domain and compared with analytical results. They are afterwards converted into time domain to characterize the scattered wavefield.\\
As shown in figure~\ref{fig:cav4dondp_valid}, we have computed the wave field around the cavity for various directions. The boundary element mesh of the cavity (2562 nodes) is generated the same way as in the previous case (Edouard {\em et al.}, 1996). This mesh has been refined since the wave field has much stronger variations compared to the previous example. In figure~\ref{fig:cav4dondp_valid}, the radial displacement $u(r,\theta,\phi,\omega)$ is displayed vs distance $\chi=r/R$ for both analytical and numerical solutions at two different normalized frequencies $\eta_P$ ($\eta_P=2R/\Lambda_P$). Different azimuthes are also considered. The analytical solution in terms of radial displacement $u_r$ is given by Pao and Mow (1973) as well as Eringen and Suhubi (1975)\footnote{There are two mistakes in the original book of Eringen and Suhubi (1975) which have to be corrected as follows. The original expression of $T_{11}^{(3)}$ in (Eringen and Suhubi, 1975) (page 914, Eq.~(9.12.11)) is:
\begin{eqnarray*}
T_{11}^{(3)}(\alpha r) = \left(n^2-n-\mbox{$\frac{1}{2}$}\beta^2r^2\right) h_n^{(1)}(\alpha r)+2\alpha r h_n^{(1)}(\alpha r)
\end{eqnarray*}
and should be replaced by the following expression:
\begin{eqnarray*}
T_{11}^{(3)}(\alpha r) = \left(n^2-n-\mbox{$\frac{1}{2}$}\beta^2r^2\right) h_n^{(1)}(\alpha r)+2\alpha r h_{n+1}^{(1)}(\alpha r)
\end{eqnarray*}
The original expression of $lC_n$ in (Eringen and Suhubi, 1975) (page 914, Eq.~(9.12.13)) is:
\begin{eqnarray*}
lC_n = (1/\Delta_n) \phi_0 i^n (2n+1) \left[ T_{11}^{(1)}(\alpha a) T_{41}^{(3)}(\alpha a) - T_{41}^{(1)}(\alpha a) T_{41}^{(3)}(\alpha a) \right]
\end{eqnarray*}
and should be replaced by:
\begin{eqnarray*}
lC_n = (1/\Delta_n) \phi_0 i^n (2n+1) \left[ T_{11}^{(1)}(\alpha a) T_{41}^{(3)}(\alpha a) - T_{41}^{(1)}(\alpha a) T_{11}^{(3)}(\alpha a) \right]
\end{eqnarray*}
where $a$ is the cavity radius, denoted $R$ in this paper.}.

\begin{figure}[htp]
\begin{center}
\end{center}
\caption{\em Diffraction of a plane wave by a spherical cavity: comparison with analytical results for various azimuthes at normalized frequencies $\eta_P$=1.00 and $\eta_P$=2.00.}
\label{fig:cav4dondp_valid}
\end{figure}

{\em Comparisons between numerical and analytical results in frequency domain.} The results are computed for various azimuthes ($\theta_i=(i-1) \times 45^0$, $1 \le i \le 5$) and figure~\ref{fig:cav4dondp_valid} displays the real part of the radial displacement vs normalized distance $\chi=r/R$ ($1 \le \chi \le 3$) at two different normalized frequencies $\eta_P$=1.00 and $\eta_P$=2.00. The analytical results are plotted with lines (dotted for $\eta_P$=1.00 and solid for $\eta_P$=2.00) and the numerical results with symbols (circles for $\eta_P$=1.00 and bullets for $\eta_P$=2.00). The agreement between the numerical and analytical results is very good for all azimuthes at $\eta_P$=1.00. For $\theta_3=90^0$, some slight differences can be noticed at $\eta_P$=2.00 near the cavity wall. This is probably due to the fact that there is a grazing incidence at this point.

{\em Scattered wavefield in time domain.} The numerical solutions are then estimated for various frequencies to compute the time domain scattered wavefield around the spherical cavity. As shown in figure~\ref{fig:cav3d_temp_v2f3}, a Ricker signal, with normalized frequency $\eta_P$=0.50, is considered for the excitation in time domain and the results are displayed for three different azimuthes: $\theta=0^o$ corresponding to the direction of propagation ($x$ axis), $\theta=45^o$ and $\theta=90^o$ ($y$ axis) that is perpendicular to the direction of propagation. For each azimuth, the time domain results are displayed for the upstream part of the propagation ($\chi \le -1$) and the downstream part ($\chi \ge +1$). For the incident and transmitted wavefields, the various azimuthes do not always coincide with the direction of propagation whereas they correspond to directions of propagation of the scattered wavefield (see following explanations). As shown in figure~\ref{fig:cav3d_temp_v2f3}, the characterization of the scattered wavefield can then be easily performed as follows:
\begin{itemize}
  \item $\theta=0^o$ (top left): for this azimuth, only the $X$ component of the displacement is displayed since the (computed) $Y$ component is found negligible. The backward and forward components of the scattered wavefield clearly appear in the figure. For the upstream part, the scattered wavefield comprises a P-wave as well as a S-wave component of respective velocities very close to the theoretical values (a few \%)\footnote{\em despite the fact we have considered a less refined mesh than for frequency domain computations.}. For the downstream part, the transmitted wavefield is easily identified and the scattered S-wave component has a velocity close to the previous value. Nevertheless, the amplitudes of the scattered wavefield components are not so large to identify them from figure~\ref{fig:cav3d_temp_v2f3}.
  \item $\theta=45^o$ (center): for this azimuth, the apparent velocity of the incident and transmitted P-waves is lower because it does not coincide with the direction of propagation. Whereas for the scattered wavefield, radial directions correspond to the direction of propagation and the time domain numerical results show a large amplitude for both $X$ and $Y$ components. For the $Y$ component of the scattered wavefield, both P and S-wave components can be identified in figure~\ref{fig:cav3d_temp_v2f3}. The velocity values estimated from the numerical results are found very close to theoretical ones. The velocity discrepancy between the downstream S component of the scattered wavefield and the transmitted P-wave is only due to the change of the apparent velocity of the latter which is azimuth dependent.
  \item $\theta=90^o$ (bottom): for this azimuth, the apparent velocity of the incident P-wave is zero because it is perpendicular to the direction of propagation. The $X$ and $Y$ components of the displacement are displayed on one side of the cavity only since they are symmetrical on the opposite side. The $X$ component clearly shows the S-wave part of the scattered wavefield. The estimation of its velocity is as good as in previous cases. For this azimuth, the $Y$ component shows that the interaction between the plane wave and the cavity is particularly complex since we have a grazing incidence on the cavity wall.
\end{itemize}

\begin{figure}[htp]
\begin{center}
\end{center}
\caption{\em Diffraction of a plane P-wave by a spherical cavity: numerical results in time domain (Ricker signal) for various azimuthes at normalized frequency $\eta_P$=0.50.}
\label{fig:cav3d_temp_v2f3}
\end{figure}

{\em Example 3}: Amplification of a plane seismic wave by a semi-spherical alluvial basin.\\
\indent{\em Description of the problem and reference solution.} The third example investigates the amplification of a plane seismic wave in an alluvial basin. In seismology and earthquake engineering, this phenomenon is known as "site effects" and generally leads to a strong amplification of the seismic motion in soft alluvial deposits (Bard and Bouchon, 1985; Bielak {\em et al.}, 1999; Chávez-García {\em et al.}, 2000; Moeen-Vaziri and Trifunac, 1985; Semblat {\em et al.}, 2000, 2003a, 2005). The example considered herein corresponds to a semi spherical alluvial basin (that is a soft elastic inclusion) in an elastic half space. Numerous papers have investigated the 3D wave diffraction by a semi spherical canyon (Lee, 1978; Liao {\em et al.}, 2004; Yokoi, 2003) or 3D seismic wave amplification by surface heterogeneities (Dravinski, 2003; Komatitsch and Vilotte, 1998; Moczo {\em et al.}, 2002; Sánchez-Sesma, 1983; Sánchez-Sesma and Luzón, 1995).

\begin{figure}[!htp]
\begin{center}
\end{center}
\caption{\em Amplification of a plane vertical P-wave by a semi-spherical basin: model description.}
\label{fig:bassin_schem}
\end{figure}

Several results have been published for the case of a semi spherical alluvial basin (Dravinski, 2003; Lee, 1984; Sánchez-Sesma, 1983). The 3D BEM model considered herein for purpose of validation is depicted in figure~\ref{fig:bassin_schem}: the mesh includes the semi-spherical basin of radius $R$ (same type of triangular meshing as in the previous section (Edouard {\em et al.}, 1996)) and part of the free-surface (for $r \le 5R$). The contribution of the free surface $r \ge 5R$ in the BIE is neglected. Therefore, the BIE are enforced at the nodes of the mesh except those located at its boundary. The model is excited by a vertical plane P-wave. For the comparison, we will consider the results of Sánchez-Sesma (1983) derived thanks to a series expansion method. We will then investigate the amplification of the motion at the surface of the alluvial basin (i.e. soft inclusion).\\

For the semi-spherical basin and the half-space, the mechanical parameters are chosen identical to Sánchez-Sesma's values as follows:
\begin{itemize}
  \item shear moduli: $\mu_R/\mu_E=0.3$
  \item mass densities: $\rho_R/\rho_E=0.6$
  \item Poisson's ratios: $\nu_R=0.30$ and $\nu_E=0.25$
\end{itemize}
where subscript $R$ refers to the alluvial basin and subscript $E$ to the half-space.\\
Similarly to the previous section, we consider for the computations the same normalized frequency as S\'anchez-Sesma corresponding to the diameter-to-wavelength ratio $\eta_P=2R/\Lambda_P$ where $\Lambda_P$ is the P wavelength in the alluvial basin.

{\em Comparison between numerical and reference results.} In figure~\ref{fig:bassin_compsesma}, the {\em amplification of the seismic motion} is computed at the free-surface (vertical displacement) and displayed vs normalized distance ($0 \le \chi \le 3$). It is compared with Sánchez-Sesma's results (1983) at normalized frequency $\eta_P=0.50$. The {\em amplification} of the vertical motion at the center of the semi-spherical basin is very well estimated by our numerical approach: 2.81 for our numerical approach (i.e. 5.63 in amplitude to be compared to 2 for the half-space) and 2.82 for Sánchez-Sesma's results (i.e. 5.64 in amplitude). The computed displacement/distance curve from our numerical approach is very close to Sán\-chez-Sesma's semi-analytical results (figure~\ref{fig:bassin_compsesma}). This amplification value is larger than for the constant depth layer case (1D) since, for the semispherical basin, focusing effects are very strong (Sánchez-Sesma, 1983; Semblat {\em et al.}, 2000, 2005).\\
It should be noticed that the normalized frequency $\eta_P=0.50$ corresponds to the fundamental frequency of the 1D case (the wavelength being $\Lambda_P=4R$ with $R$ the depth of the basin). Nevertheless, at this frequency, the variation of the amplification factor vs frequency is strong: for the 3D semi-spherical basin, this frequency is rather far from the maximum amplification peak. If we compute the amplification factor at the centre of the semi-spherical basin for various frequencies, the largest site effects are found at normalized frequency $\eta_P=0.57$. At this frequency and for the mechanical properties chosen herein, the corresponding {\em amplification factor} is about 4.76 (i.e. 9.52 in amplitude), that is 70\% larger than for $\eta_P=0.50$. For sake of comparisons, around normalized frequency $\eta_P=0.57$ the amplitude variation with frequency is smaller (resp. basin properties).\\

\begin{figure}[htp]
\begin{center}
\end{center}
\caption{\em Computed vertical motion showing the amplification at the basin surface and comparison with Sánchez-Sesma's result for normalized frequency $\eta_P=0.50$.}
\label{fig:bassin_compsesma}
\end{figure}

\vspace{1cm}
\begin{center}
{\large{Conclusion}}
\end{center}

In this paper, a simple and efficient method to regularize singular integrals in 3D boundary integral equations has been presented. The regularization me\-thod is based on the derivation of analytical terms which correct the error due to the straightforward estimation of singular integrals through classical Gaussian quadrature formulas. The analytical derivation of the correction term has assumed a three-noded triangle discretization of the boundary surfaces. However, the method described in the appendix can be easily generalized to any flat element such as quadrangle. This method has been implemented in a BEM code and applied to 3D frequency domain elastodynamics.\\
Some comparisons have been made with (semi-)ana\-ly\-tical results for simple problems:
\begin{itemize}
  \item {\em cavity under harmonic pressure}: the agreement between our numerical results and the analytical solution is very good for various frequencies even with a small number of nodes/elements. The efficiency of the regularization method proposed in this paper is also discussed for this example.
  \item {\em diffraction of a plane-wave by a spherical cavity}: the agreement between our numerical results and the analytical solution is very good for various azimuthes and frequencies. In time domain, the numerical results are also found to be satisfactory.
  \item {\em wave amplification in a semispherical alluvial basin (soft inclusion)}: the comparison of our numerical results with Sánchez-Sesma's semi-analytical results (Sánchez-Sesma, 1983) is also satisfactory. Further comparisons are planned with other current numerical approaches and more complex geometries.
\end{itemize}
Considering these good results, future work will then concern more realistic cases in the field of seismology. For sake of numerical efficiency, the regularization me\-thod could also be implemented in a Symmetric Galerkin boundary element formulation (Bonnet {\em et al.}, 1998) or in the framework of a Fast Multipole Method (Greengard {\em et al.}, 1998). Our main goal is to have a detailed description of the 3D geological structure of a given area to perform reliable computations of seismic wave propagation and amplification (Bard and Bouchon, 1985; Bouchon {\em et al.}, 1989; Chávez-García {\em et al.}, 2000; Frankel and Vidale, 1992; Moczo {\em et al.}, 2002; Semblat {\em et al.}, 2000, 2003a,b, 2005).

\newpage

\appendix

\begin{center}
{\large{Appendix}}
\end{center}

Calculation of  $C_i^k(\u{y};E_e)$\\
\indent Let us calculate the free term $C_i^k(\u{y})$ defined by:
\begin{equation}
C_i^k(\u{y})=\lim_{\varepsilon\rightarrow 0} \int_{s_\varepsilon} T_i^k(\u{x},\u{y})\,dS_x
\label{eq:cij_0}
\end{equation}
In Eq. (\ref{eq:cij_0}), $s_\varepsilon$ is a spherical surface of radius $\varepsilon$. Let $\u{x}$ be a point on $s_\varepsilon$. The unit outward normal to $s_\varepsilon$ is given by $\u{n} = (\u{y} - \u{x})/\varepsilon$. Thus the stress vector of the Kelvin fundamental solution applied to $s_\varepsilon$ has the form:
\begin{equation}
T_i^k(\u{x},\u{y}) = \frac{1}{8\pi(1-\nu)}\frac{1}{\varepsilon^2} \left ( (1-2\nu)\delta_{ik} + 3n_in_k \right )
\end{equation} 
Substituting this expression for $T_i^k$ in (\ref{eq:cij_0}) yields:
\begin{eqnarray}\label{eq:cik_1}
C_i^k(\u{y}) & = & \frac{1}{8\pi(1-\nu)}.\\
& & \lim_{\varepsilon\rightarrow 0} \frac{1}{\varepsilon^2} \left( (1-2\nu)|s_\varepsilon|\delta_{ik}
+3\int_{s_\varepsilon}n_in_k\,dS_x \right)
\nonumber
\end{eqnarray}
where $|s_\varepsilon|$ is the surface area of $s_\varepsilon$. Here $|s_\varepsilon|=\varepsilon^2\psi$, where $\psi$ is the solid angle. A small amount of calculations allows to derive the following expression:
\begin{equation}
\int_{s_\varepsilon}n_in_k\,dS_x = \frac{|s_\varepsilon|}{3}\delta_{ik} - \frac{2}{3}\varepsilon^2 \!\!\sum_{e \in I(\u{y})}\sin\left(\frac{\theta^e}{2}\right)b_i^en_k^e
\label{eq:n_in_k}
\end{equation}
where $\theta^e$ is the angle formed by the edges of $E_e$ at $\u{y}$, $b_i^e$ is the unit vector of the bissecting line and $n_k^e$ is the unit outward normal to $E_e$. The symmetry with respect to subscripts $i$ and $k$ in Eq. (\ref{eq:n_in_k}) shows the following identity:
\begin{equation}
\sum_{e \in I(\u{y})}\sin\left(\frac{\theta^e}{2}\right)b_i^en_k^e = \!\!\sum_{e \in I(\u{y})}\sin\left(\frac{\theta^e}{2}\right)b_k^en_i^e
\label{eq:sym_ik}
\end{equation}
Combining (\ref{eq:cik_1}) and (\ref{eq:n_in_k}) yields:
\begin{eqnarray}\label{eq:cik_2}
C_i^k(\u{y}) & = & \frac{\psi}{4\pi}\delta_{ik}\\
& & - \frac{1}{8\pi(1-\nu)} \!\!\sum_{e \in I(\u{y})}\sin\left(\frac{\theta^e}{2}\right)\left ( b_i^en_k^e +  b_k^en_i^e \right )
\nonumber
\end{eqnarray}
Finally the solid angle is assumed to be the sum of element solid angles:
\begin{equation}
\psi = \!\!\sum_{e \in I(\u{y})} \psi^e
\label{eq:psi}
\end{equation}
Pratically, as shown in figure~\ref{fig:coin}, $\psi^e$ can be defined by the solid angle of the trihedron  of apex $\u{y}$ formed by the two edges of element $E_e$ and the semi-axis in the direction of $-\u{n}(\u{y})$, where $\u{n}(\u{y})$ is an arbitrary outward unit vector at $\u{y}$. In this case $\psi^e=(\varphi^e_1+\varphi^e_2+\varphi^e_3 - \pi)$ where the $\varphi^e_i$ are the three angles formed by the plane of the trihedron (figure \ref{fig:fcij-2ab}). The calculation of the solid angles $\psi_e$ relies on the knowledge of an outward unit vector at each node of the mesh. Practically for each node of coordinate $\u{y}$, $\u{n}(\u{y})$ can be calculated as the mean of unit normals to each element of $I(\u{y})$. It should be noticed that the averaging of the normal is only conventional. It results from an arbitrary choice in order to perform the calculation of solid angles $\psi_e$. The accuracy of the method does not depend on this averaging procedure since the value of the solid angle $\psi$ (equation~(\ref{eq:psi})) is eventually recovered whatever the choice of $\u{n}(\u{y})$. Therefore, the free term $C_i^k(\u{y})$ is really the sum of free term elements $C_i^k(\u{y};E_e)$ of the form:
\begin{eqnarray}
C_i^k(\u{y};E_e) & = & \frac{\psi^e}{4\pi}\delta_{ik}\label{eq:cij_7}\\
& & - \frac{1}{8\pi(1-\nu)} \sin\left(\frac{\theta^e}{2}\right)\left ( b_i^en_k^e +  b_k^en_i^e \right)
\nonumber
\end{eqnarray}

\begin{figure}[htp]
\begin{center}
\end{center}
\caption{\em Conical surface with apex at $\u{y}$}
\label{fig:coin}
\end{figure}

\begin{figure}[htp]
\begin{center}
\end{center}
\caption{\em Description of angles $\varphi_1$, $\varphi_2$ and $\varphi_3$ defining the element solid angle $\psi^e$.}
\label{fig:fcij-2ab}
\end{figure}

Calculation of $I_i^k(\u{y};E_e)$\\
\indent The integral $I_i^k(\u{y})$ can be written in the form:
\begin{equation}
I_i^k(\u{y}) = \lim_{\varepsilon\rightarrow 0} \!\!\sum_{e \in I(\u{y})} \int_{(E_e-e_{\varepsilon}^e)} T_i^k(\u{x},\u{y}) dS_x
\label{eq:I1_1}
\end{equation}
where $e_{\varepsilon}^e = E_e \cap v_\varepsilon$ (with of course $\sum_{e \in I(\u{y})} e_{\varepsilon}^e = e_{\varepsilon}$). Given $\varepsilon > 0$, let us calculate the element integral appearing in (\ref{eq:I1_1}). Let $\u{x}$ be a current point on $E_e$ and note $n_i^e$ the unit outward normal to $E_e$. The stress vector of the Kelvin fundamental solution applied to $E_e$ is given by:
\begin{equation}
T_i^k(\u{x},\u{y}) = \frac{1-2\nu}{8\pi(1-\nu)} \frac{n_i^ee_k - n_k^ee_i}{r^2}
\end{equation}
where $e_i = (y_i - x_i)/r$. A trivial integration of the above expression shows that:
\begin{eqnarray}\nonumber
\int_{(E_e-e_{\varepsilon}^e)} \frac{n_i^ee_k - n_k^ee_i}{r^2} dS_x =\\
\int_0^{\theta^e}(n_i^ee_k(\alpha) - n_k^ee_i(\alpha)) \ln L(\alpha) d\alpha
\label{eq:I1_2}\\
-2\sin\left(\frac{\theta^e}{2}\right)(n_i^eb_k^e - n_k^eb_i^e) \ln\varepsilon
\nonumber
\end{eqnarray}
where $L(\alpha)$ is the length of the segment defined in the figure (\ref{fig:lalpha}). Thanks to identity (\ref{eq:sym_ik}), integral $I_i^k(\u{y})$ is then the sum of element integrals defined by:
\begin{eqnarray}
I_i^k(\u{y};E_e) =~~~~~~~~~~~~~~~~~~~~~~~~~~~~~~~~~~~~~~~~~~~~~~\label{I2}\\
\nonumber\frac{1-2\nu}{8\pi(1-\nu)} \int_0^{\theta^e}(n_i^ee_k(\alpha) - n_k^ee_i(\alpha)) \ln L(\alpha) d\alpha
\end{eqnarray}

\begin{figure}[htp]
\begin{center}
\end{center}
\caption{\em Distance $L(\alpha)$ on $E_e$}
\label{fig:lalpha}
\end{figure}

Calculation of $J_i^k(\u{y};E_e)$\\
\indent The integral $J_i^k(\u{y})$ can be written in the form:
\begin{equation}
J_i^k(\u{y}) = \lim_{\varepsilon\rightarrow 0} \!\!\sum_{e \in I(\u{y})} \int_{(E_e-e_{\varepsilon}^e)} U_i^k(\u{x},\u{y}) dS_x
\label{eq:j1_1}
\end{equation}
Substituting expression (\ref{eq:Uik_kelvin}) for $U_i^k(\u{x},\u{y})$ in (\ref{eq:j1_1}) gives the expression of $J_i^k(\u{y};E_e)$:
\begin{eqnarray}
J_i^k(\u{x};E_e) =~~~~~~~~~~~~~~~~~~~~~~~~~~~~~~~~~~~~~~~~~~~~~\label{eq:j1_2}\\
\nonumber\frac{1}{16\pi\mu(1-\nu)} \int_{0}^{\theta^e} (e_i e_k + (3-4\nu)\delta_{ik}) L(\alpha) d\alpha
\end{eqnarray}

\newpage


\begin{center}
{\large{References}}
\end{center}

\begin{small}

\begin{description}
\item Aki, K. and K.L. Larner (1970).
\newblock Surface Motion of a Layered Medium Having an Irregular Interface due to Incident Plane SH Waves,
\newblock {\em J. Geophys. Res.}, {\bf 75} 1921-1941.

\item Aubry, D., and D. Clouteau (1991).
\newblock A regularized boundary element method for stratified media,
\newblock {\em 1st Int. Conf. on Mathematical and Numerical Aspects of Wave Propagation, SIAM}.

\item Banerjee, P.K., S. Ahmad, and K. Chen (1988).
\newblock Advanced application of BEM to wave barriers in multi-layered three-dimensional soil media,
\newblock {\em Earthquake Eng. Struct. Dyn.} {\bf 16}, 1041-1060.

\item Bard, P.Y., and M. Bouchon (1985).
\newblock The two dimensional resonance of sediment filled valleys, 
\newblock {\em Bull. Seism. Soc. Am.} {\bf 75}, 519-541.

\item Beskos, D.E. (1997).
\newblock Boundary elements methods in dynamic analysis: Part II (1986-1996),
\newblock {\em Appl. Mech. Rev. (ASME)} {\bf 50}(3), 149-197.

\item Beskos, D.E., B. Dasgupta, and I.G. Vardoulakis (1986).
\newblock Vibration isolation using open or filled trenches,
\newblock {\em Computational Mech.} {\bf 1}, 43-63.

\item Bielak, J., J. Xu, and O. Ghattas (1999).
\newblock Earthquake ground motion and structural response in alluvial valleys,
\newblock {\em J. Geotech. Geoenviron. Eng.} {\bf 125}, 413-423.

\item Bonilla, F. (2000).
\newblock {\em Computation of Linear and Nonlinear Site Response for Near Field Ground Motion},
\newblock Ph.D., University of California at Santa Barbara.

\item Bonnet, M. (1999).
\newblock {\em Boundary integral equation methods for solids and fluids}.
\newblock Wiley.

\item Bonnet, M., and H.D. Bui (1993).
\newblock Regularization of the displacement and traction BIE for 3D elastodynamics using indirect methods.
\newblock in {\em Advances in Boundary Element Techniques}, Kane JH, Maier G., Tosaka N. and Atluri SN (eds), Springer-Verlag, Berlin, 1-29.

\item Bonnet, M., and H.H. Xiao (1995)
\newblock Computation of energy release rate using material differentiation of elastic BIE for 3-D elastic fracture.
\newblock {\em Eng. Analysis Boundary Elements} {\bf 15}, 137-149.

\item Bonnet, M., G. Maier, and C. Polizzotto (1998).
\newblock Symmetric Galerkin boundary element methods.
\newblock {\em ASME Appl. Mech. Rev.} {\bf 51}(11), 669-704.

\item Bouchon, M., M. Campillo, and S. Gaffet (1989).
\newblock A Boundary Integral Equation-Discrete Wavenumber Representation Method to Study Wave Propagation in Multilayered Media Having Irregular Interfaces,
\newblock {\em Geophysics} {\bf 54}, 1134-1140.

\item Bui, H.D., B. Loret, and M. Bonnet (1985).
\newblock Régularisation des équations intégrales de
l'élastostatique et de l'élastodynamique.
\newblock {\em C.R.Acad. Sci. Paris, série II} {\bf 300}, 633-636.

\item Chammas, R., O. Abraham, P. Cote, H.A. Pedersen, and J.F. Semblat (2003).
\newblock Characterization of heterogeneous soils using surface waves: homogenization and numerical modeling.
\newblock {\em Int. J. Geomech. (ASCE)} {\bf 3}(1), 55-63.

\item Chávez-García, F.J., D.G. Raptakis, K. Makra, and K.D. Pitilakis (2000).
\newblock Site effects at Euroseistest-II. Results from 2D numerical modelling and comparison with observations.
\newblock {\em Soil Dyn. Earthquake Eng.} {\bf 19}(1), 23-39.

\item Chen, J.T., and H.K. Hong (1999)
\newblock Review of dual boundary element methods with emphasis on hypersingular integrals and divergent series.
\newblock {\em Appl. Mech. Rev.} {\bf 52}(1), 17-33.

\item Dangla, P. (1988).
\newblock A plane strain soil-structure interraction model.
\newblock {\em Earthquake Eng. Struct. Dyn.} {\bf 16}, 1115-1128.

\item Dangla, P. (1990).
\newblock Couplage éléments finis-équations intégrales en élastodynamique et interaction sol-structure.
\newblock {\em Rapport Études et Recherches (Série: Mécanique et Mathématiques Appliquées)}, Laboratoire Central des Ponts et Chaussées, Paris.

\item Dravinski, M. (2003).
\newblock Scattering of elastic waves by a general anisotropic basin. Part 2: a 3D model,
\newblock {\em Earthquake Eng. Struct. Dyn.} {\bf 32}, 653-670.

\item Edouard, S., B. Legras, F. Lefevre, and R. Eymard (1996).
\newblock The effect of small-scale inhomogeneities on ozone depletion in the Arctic,
\newblock {\em Nature} {\bf 384}.

\item Eringen, A.C., and E.S. Suhubi (1975).
\newblock {\em Elastodynamics (vol. II-linear theory)}. Academic Press.

\item Faccioli, E., F. Maggio, A. Quarteroni, and A. Tagliani (1996).
\newblock Spectral Domain Decomposition Methods for the Solution of Acoustic and Elastic Wave Equations,
\newblock {\em Geophysics} {\bf 61}, 1160-1174.

\item Frankel, A. and J. Vidale (1992).
\newblock A Three-Dimensional Simulation of Seismic Waves in the Santa Clara Valley, California, from a Loma Prieta Aftershock,
\newblock {Bull. Seism. Soc. Am.} {\bf 82}, 2045-2074.

\item Greengard, L., J. Huang, V. Rokhlin, and S. Wandzura (1998).
\newblock Accelerating fast multipole methods for the Helmholtz equation at low frequencies.
\newblock {\em IEEE Comp. Sci. Eng.} {\bf 5}(3), 32-38.

\item Guiggiani, M., Gigante, M. (1990)
\newblock A general algorithm for multidimensional Cauchy principal value integrals in the boundary element method.
\newblock {\em ASME J. Appl. Mech.} {\bf 57}, 906-915.

\item Guiggiani, M., Krishnasamy, G., Rudolphi, T. J., Rizzo, F. J. (1992)
\newblock A general algorithm for the numerical solution of hypersingular boundary integral equations.
\newblock {\em ASME J. Appl. Mech.} {\bf 59}, 604-614.

\item Guiggiani, M. (1994).
\newblock Hypersingular formulation for boundary stress evaluation.
\newblock {\em Eng. Analysis Boundary Elements.} {\bf 13}, 169-179.

\item Humbert, P., G. Fezans, A. Dubouchet, and D. Remaud (2005).
\newblock CESAR-LCPC~: a modeling software package dedicated to civil engineering,
\newblock {\em Bull. des Laboratoires des Ponts et Chauss\'ees} {\bf 256-257} (to be published).

\item Ihlenburg, F., and I. Babu\v{s}ka (1995)
\newblock Dispersion analysis and error estimation of Galerkin finite element methods for the Helmholtz equation. \newblock {\em Int. J. for Numerical Methods in Eng.} {\bf 38}, 3745-3774.

\item Komatitsch, D., and J.P. Vilotte (1998).
\newblock The Spectral Element Method: An Efficient Tool to Simulate the Seismic Response of 2D and 3D Geological Structures,
\newblock {\em Bull. Seism. Soc. Am.} {\bf 88}, 368-392.

\item Lachat, J.G., and J.O. Watson (1976).
\newblock Effective numerical treatment of boundary integral equations: a formulation for three-dimensional elastostatics.
\newblock {\em Int. J. Num. Meth. Engng.} {\bf 10}, 991-1005.

\item Lee, V.W. (1978).
\newblock Displacements near a three-dimensional hemispherical canyon subjected to incident plane waves,
\newblock {\em Report No. CE 78-16, University of Southern California}, 126 p.

\item Lee, V.W. (1984).
\newblock Three-dimensional diffraction of plane P, SV \& SH waves by a hemispherical alluvial valley,
\newblock {\em Soil Dyn. Earthquake Eng.} {\bf 3}, 133-144.

\item Liao, W.I., T.J. Teng, and C.S. Yeh (2004).
\newblock A series solution and numerical technique for wave diffraction by a three-dimensional canyon,
\newblock {\em Wave Motion} {\bf 39}, 129-142.

Manolis, G.D., and D.E. Beskos (1988).
\newblock {\em Boundary Element Methods in Elastodynamics}.
\newblock Unwin-Hyman (Chapman \& Hall), London.

\item Moczo, P., J. Kristek, V. Vavrycuk, R.J. Archuleta, and L. Halada (2002).
\newblock 3D Heterogeneous Staggered-Grid Finite-Difference Modeling of Seismic Motion with Volume Harmonic and Arithmetic Averaging of Elastic Moduli and Densities,
\newblock {\em Bull. Seism. Soc. Am.} {\bf 92}(8), 3042-3066. 

\item Moeen-Vaziri, N., and M.D. Trifunac (1985).
\newblock Scattering of plane SH-waves by cylindrical canals of arbitrary shape,
\newblock {\em Soil Dyn. Earthquake Eng.} {\bf 4}(1), 18-23.

\item Niu, Z.R., and H.L. Zhou (2004).
\newblock The natural boundary integral equation in potential problems and regularization of the hypersingular integral.
\newblock {\em Computers and Structures} {\bf 82}, 315-323.

\item Pao, Y.H., and C.C. Mow (1973).
\newblock {\em Diffraction of elastic wa\-ves and dynamic stress concentrations.}
\newblock Crane, Russak \& Company Inc.

\item Pecker, A. (1984).
\newblock {\em Soil dynamics}.
\newblock Paris, Presses de l'Ecole Nationale des Ponts et Chaussées, 259pp. (in French).

\item Sánchez-Sesma, F.J. (1983).
\newblock Diffraction of elastic waves by three-dimensional surface irregularities,
\newblock {\em Bull. Seism. Soc. Am.} {\bf 73}(6), 1621-1636.

\item Sánchez-Sesma, F.J., and F. Luzón (1995).
\newblock Seismic Response of Three-Dimensional Alluvial Valleys for Incident P, S and Rayleigh Waves,
\newblock {\em Bull. Seism. Soc. Am.} {\bf 85}, 269-284.

\item Semblat, J.F., and M.P. Luong (1998).
\newblock Wave propagation through soils in centrifuge experiments.
\newblock {\em J. Earthquake Eng.} {\bf 2}(1), 147-171.

\item Semblat, J.F., and J.J. Brioist (2000).
\newblock Efficiency of higher order finite elements for the analysis of seismic wave propagation.
\newblock {\em J. Sound Vib.} {\bf 231}(2), 460-467.

\item Semblat, J.F., A.M. Duval, and P. Dangla (2000).
\newblock Numerical analysis of seismic wave amplification in Nice (France) and comparison with experiments.
\newblock {\em Soil Dyn. Earthquake Eng.} {\bf 19}(5), 347-362.

\item Semblat, J.F., A.M. Duval, and P. Dangla (2002).
\newblock Seismic site effects in a deep alluvial basin: numerical analysis by the boundary element method,
\newblock {\em Comput. Geotech.} {\bf 29}(7), 573-585.

\item Semblat, J.F., A.M. Duval, and P. Dangla (2003a).
\newblock Modal Superposition Method for the Analysis of Seismic-Wave Amplification.
\newblock {\em Bull. Seism. Soc. Am.} {\bf 93}(3), 1144-1153.

\item Semblat, J.F., Paolucci R., Duval A.M. (2003b).
\newblock Simplified vibratory characterization of alluvial basins 
\newblock {\em C. R. Geoscience} {\bf 335}, 365-370.

\item Semblat, J.F., M. Kham, E. Parara, P.Y. Bard, K. Pitilakis, K. Makra, and D. Raptakis (2005).
\newblock Seismic wave amplification: basin geometry vs soil layering. 
\newblock {\em Soil Dyn. Earthquake Eng.} (to be published).

\item Sladek, V., and J. Sladek (1998).
\newblock Singular integrals and boundary elements.
\newblock {\em Comput. Methods Appl. Mech. Eng.} {\bf 157}, 251-266.

\item Sladek, V., and J. Sladek (1996).
\newblock Regularization of hypersingular integrals in BEM formulations using various kinds of continuous elements.
\newblock {\em Eng. Anal. Bound. Elem.} {\bf 17}, 5-18.

\item Sladek, V., J. Sladek, and M. Tanaka (2001).
\newblock Numerical integration of logarithmic and nearly logarithmic singularity in BEMs.
\newblock {\em Applied Mathematical Modelling} {\bf 25}, 901-922.

\item Sladek, V., J. Sladek, and M. Tanaka (1997).
\newblock Evaluation of $1/r$ integrals in BEM formulations for 3-D problems using coordinate multitransformations.
\newblock {\em Eng. Analysis with Boundary Elements} {\bf 20}, 229-244.

\item Tanaka, M., V. Sladek, and J. Sladek (1994).
\newblock Regularization Techniques Applied to Boundary Element Methods.
\newblock {\em Appl. Mech. Rev.} {\bf 47}, 457-499.

\item Virieux, J. (1986).
\newblock P-SV Wave Propagation in Heterogeneous Media: Velocity-Stress Finite-Difference Method,
\newblock {\em Geophysics} {\bf 51}, 889-901.

\item Xiao, H.H. (1994).
\newblock {\em Équations intégrales de frontière, dérivation par rapport au domaine et approche énergétique pour les solides élastiques fissurés}.
\newblock PhD thesis, École Polytechnique, Palaiseau, France.

\item Yokoi, T. (2003).
\newblock The higher order Born approximation applied to improve the solution of seismic response of a three-dimensional canyon by the Indirect Boundary Method.
\newblock {\em Physics of the Earth and Planetary Interior}, {\bf 137}, 97-106.

\end{description}

\end{small}

\end{twocolumn}


\end{document}